\documentclass{article}

\usepackage{PRIMEarxiv}

\usepackage[utf8]{inputenc} 
\usepackage[T1]{fontenc}    
\usepackage{hyperref}       
\usepackage{url}            
\usepackage{booktabs}       
\usepackage{amsfonts}       
\usepackage{nicefrac}       
\usepackage{microtype}      
\usepackage{lipsum}
\usepackage{fancyhdr}       
\usepackage{graphicx}       
\graphicspath{{media/}}     
\usepackage{enumitem}
\usepackage{caption}
\usepackage{subcaption}
\usepackage{amsthm}

\title{Are you aware of what you are watching? Role of machine heuristic in online content recommendations
}

\author{
  Soyoung Oh, Eunil Park \\
  Applied Artificial Intelligence \\
  Sungkyunkwan University \\
  Seoul, Republic of Korea\\
  \texttt{\{sori424, eunilpark\}@g.skku.edu@skku.edu} \\
}

\begin{document}
\maketitle

\begin{abstract}
Since recommender systems have been created and developed to automate the recommendation process, users can easily consume their desired video content on online platforms. In this line, several content recommendation algorithms are introduced and employed to allow users to encounter content of their interests, effectively. However, the recommendation systems sometimes regrettably recommend inappropriate content, including misinformation or fake news. To make it worse, people would unreservedly accept such content due to their cognitive heuristic, machine heuristic, which is the rule of thumb that machines are more accurate and trustworthy than humans. In this study, we designed and conducted a web-based experiment where the participants are invoked machine heuristic by experiencing the whole process of machine or human recommendation system. The results demonstrated that participants (N = 89) showed a more positive attitude toward a machine recommender than a human recommender, even the recommended videos contain inappropriate content. While participants who have a high level of trust in machines exhibited a negative attitude toward recommendations. Based on these results, we suggest that a phenomenon known as algorithm aversion might be simultaneously considered with machine heuristic in investigating human interaction with a machine.  
\end{abstract}


\section{Introduction}
    As online content recommendation systems have penetrated deeply into the daily use of online technology, service providers have made significant efforts to provide appropriate content that meets users’ preferences. One solution involves employing a unique and robust algorithm to address the user’s needs via system-initiated personalization. For instance, YouTube, the world’s most popular online video-sharing platform, provides the auto-play button which plays another video once the current video is finished by predicting what the users would like to see based on their viewing history~\cite{bedjaoui2018user}. However, the convenience of such automation makes it inevitable for users to face several problems~\cite{pariser2011filter, tufekci2018youtube}. One of the proposed issues is the ``filter bubble'' which results in biased recommendations on specific political opinions~\cite{nie2014social}. Owing to their customization, users would be exposed to the content they favor resulting in narrowing users' chances of exploring other opinions~\cite {pariser2011filter}. For example, in online media channels, the news algorithm overrepresents right-wing voices which is characterized by hateful vocabulary, violent content, and discriminatory biases, while under-represents others~\cite{ottoni2018analyzing}. Moreover, the ``autoplay'' button, which restricts users to perform any additional activities to watch the next video, can create an infinite loophole of presenting more biased videos~\cite{tufekci2018youtube}. For instance, videos about vegetarianism led to videos about veganism, and videos about jogging lead to videos about running ultra-marathons. Such a problem is related to the revenue-making system of these platforms. As higher views produce higher revenue, it is inevitable for the platforms to recommend content that users would like to increase viewership in the platforms~\cite{flederblockbuster, hosanagar2014will}.

    However, such extreme recommendations can prohibit individuals from being adequately informed about the status quo and rational thinking~\cite{haim2018burst}. Moreover, these problems can be accelerated with algorithmic cues of the recommendation systems. As humans are known as cognitive misers who tend to make decisions on the basis of cognitive heuristics, which can be triggered by contextual cues in the situation or on the interface~\cite{sundar2008main}. Along with the machine-like cues in the recommendation systems, the machine heuristic, the rule of thumb that machines are more objective and trustworthy than humans~\cite{sundar2019machine}, can contribute to creating a positive attitude toward the algorithmic recommendations. That is, the interface cues trigger the humans' instinctive preference for machine agents over human agents to lead them unconsciously consume the recommended videos without noticing inappropriate contents within the recommendations. In this paper, we aim to test this probability by answering the question (RQ):
    
    \begin{itemize}
        \item \textbf{RQ}: Are there notable differences in users' attitudes toward machine-recommended and human-recommended videos?
    \end{itemize}

    Moreover, considering the prior studies on the notable roles of source credibility~\cite{stroud2013perceptions, wolker2018algorithms}, we assumed that credibility which is led by recommender cue would positively affect the attitude toward recommended videos. Thus, the following hypothesis is proposed:

    \begin{itemize}
        \item \textbf{H1}: When people find the recommendation system more credible, depending on whether the video is machine-recommended or human-recommended, they have a more positive attitude towards recommendations.
    \end{itemize}
    
    Trust is a critical factor that impacts not only face-to-face interpersonal relationship, but also important for human-machine communications~\cite{yu2018trust, hoff2015trust, lee1994trust}. Particularly for the systems where users are required to make decisions based, at least partially, on machine recommendations. The experiential effects based on the context-dependent variables in machine system can directly influence the trust~\cite{cohen1997trust}. For example, the type of system, complexity, and the difficulty of the task determines the trust in automated system. In addition to the external factors, as humans are creatures of experience, they can use pre-existing knowledge from past interactions with automation when assessing the trustworthiness of novel systems~\cite{hoff2015trust}. The initially learned trust built on prior experience with the system or similar technology~\cite{yuviler2011effect} and preexisting attitudes as well as expectations~\cite{abe2006alarm} can alter the formation process of the trust and subsequent decisions during interacting with automated systems. As machine heuristic is primed by the cognitive accessibility at the time of decision making~\cite{higgins1985nature}, the users' predispositions of the trust in the source of the recommender would interact with the recommender cues. In other words, the effect of reliance on a machine-like recommendation cue is dependent upon prior experiences with machines. However, this effect will not be found when the human recommender cue is presented. Thus, we propose the following hypothesis:

    \begin{itemize}
        \item \textbf{H2}: The high level of predispositions of trust in machines makes people have a positive attitude to machine-recommended videos.  
    \end{itemize}

\begin{figure}[h]
        \begin{center}
            \includegraphics[width=0.6\linewidth]{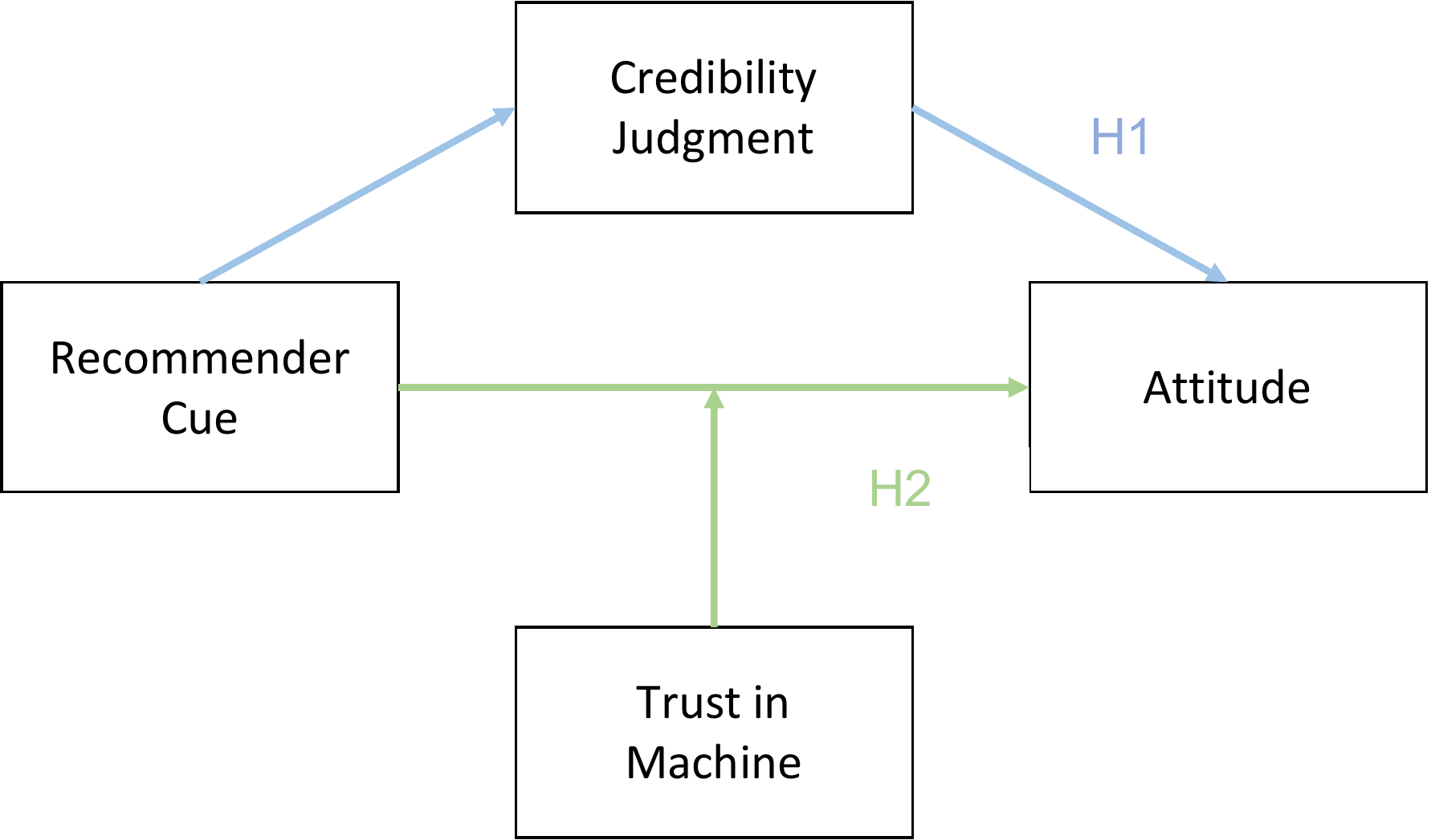}
            \caption{A proposed research model to analyze machine heuristic in recommendation system}
            \label{fig:research_model}
        \end{center}
    \end{figure}

    Based on the research question and suggested hypotheses, the research model in Figure~\ref{fig:research_model} is developed to illustrate the role of machine heuristic in credibility judgment and attitude.

\section{Experiment}

\subsection{Method}
    Our study received the institutional review board (IRB) approval from the affiliated university (2020-11-025). A between-subjects experiment was conducted to validate the hypotheses and the research model. Participants experienced to one of the recommendation systems (i.e., human versus machine). They were then instructed to respond to a survey, measuring users' degree of trust in the machine, credibility of the system, and attitude toward the recommended video list which consists of appropriate and inappropriate contents. Our experimental stimulus can be replicated by using publicly available codes\footnote{\url{https://github.com/sori424/Machine-Heuristic-Experimental-Stimulus}}. 
    
\subsection{Participants}
    A total of 90 participants were recruited from the South Korean professional survey agency. After a filtering procedure, 89 validated responses were organized by 35 males (39.3\%) and 54 females (60.7\%). The age ranged from 20 to 60 years. While 43.8\% were in their 20s, 32.6\% were in their 30s, 15.7\% were in their 40s, and the rest were in their 50s (7.9\%). The educational background ranged from below high school graduate to Ph.D., and the majority of the participants had a bachelor's degree (60.7\%). Half of the participants were assigned to the machine recommender condition and the others were assigned to the human recommender condition.

\subsection{Experimental Stimulus and Procedure}
\begin{figure*}
  \centering
    \begin{subfigure}[b]{0.45\textwidth}
         \centering
         \includegraphics[width=\textwidth]{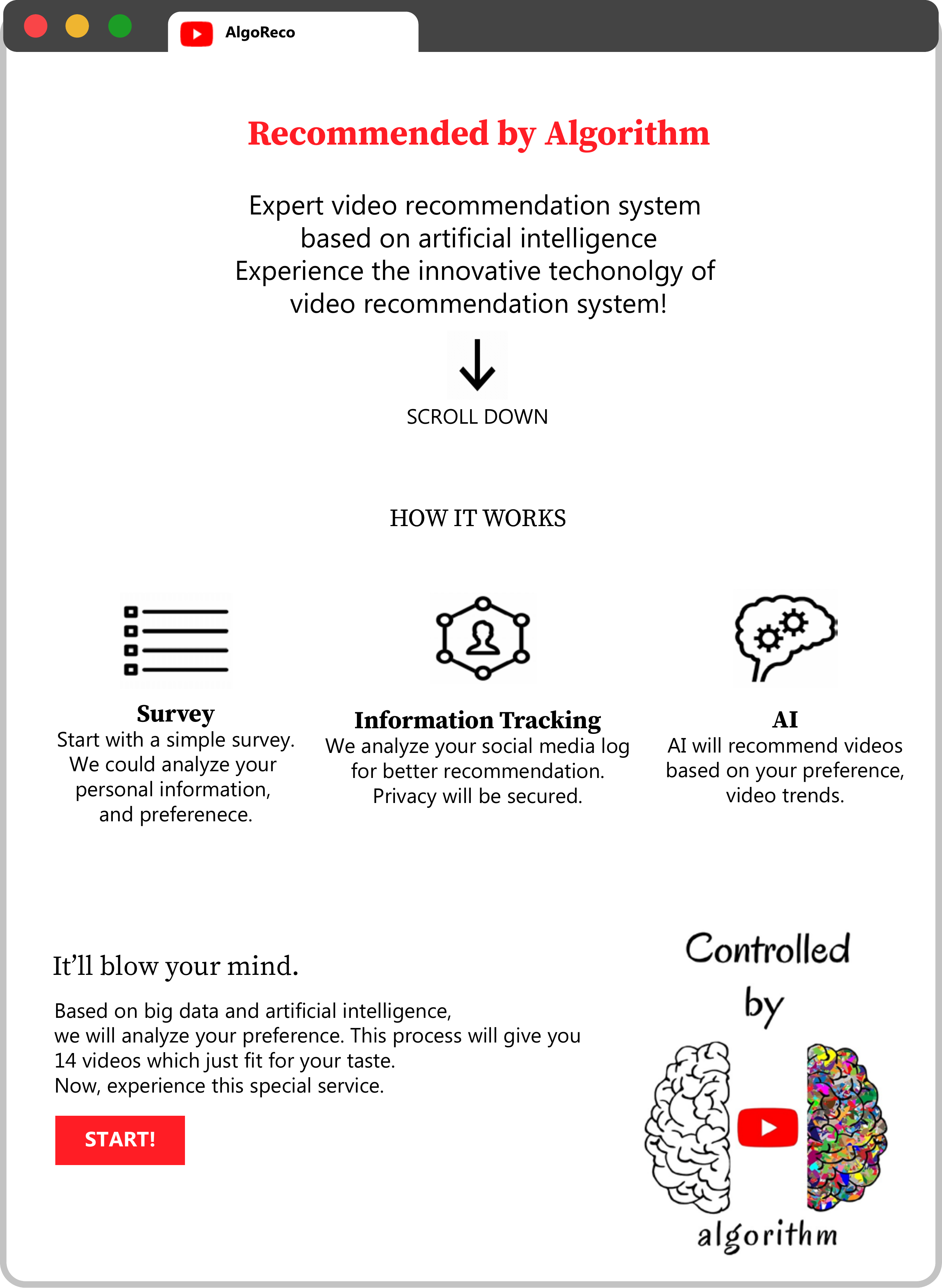}
         \caption{Machine condition}
         \label{fig:y equals x}
     \end{subfigure}
    \quad\quad
    \begin{subfigure}[b]{0.45\textwidth}
         \centering
         \includegraphics[width=\textwidth]{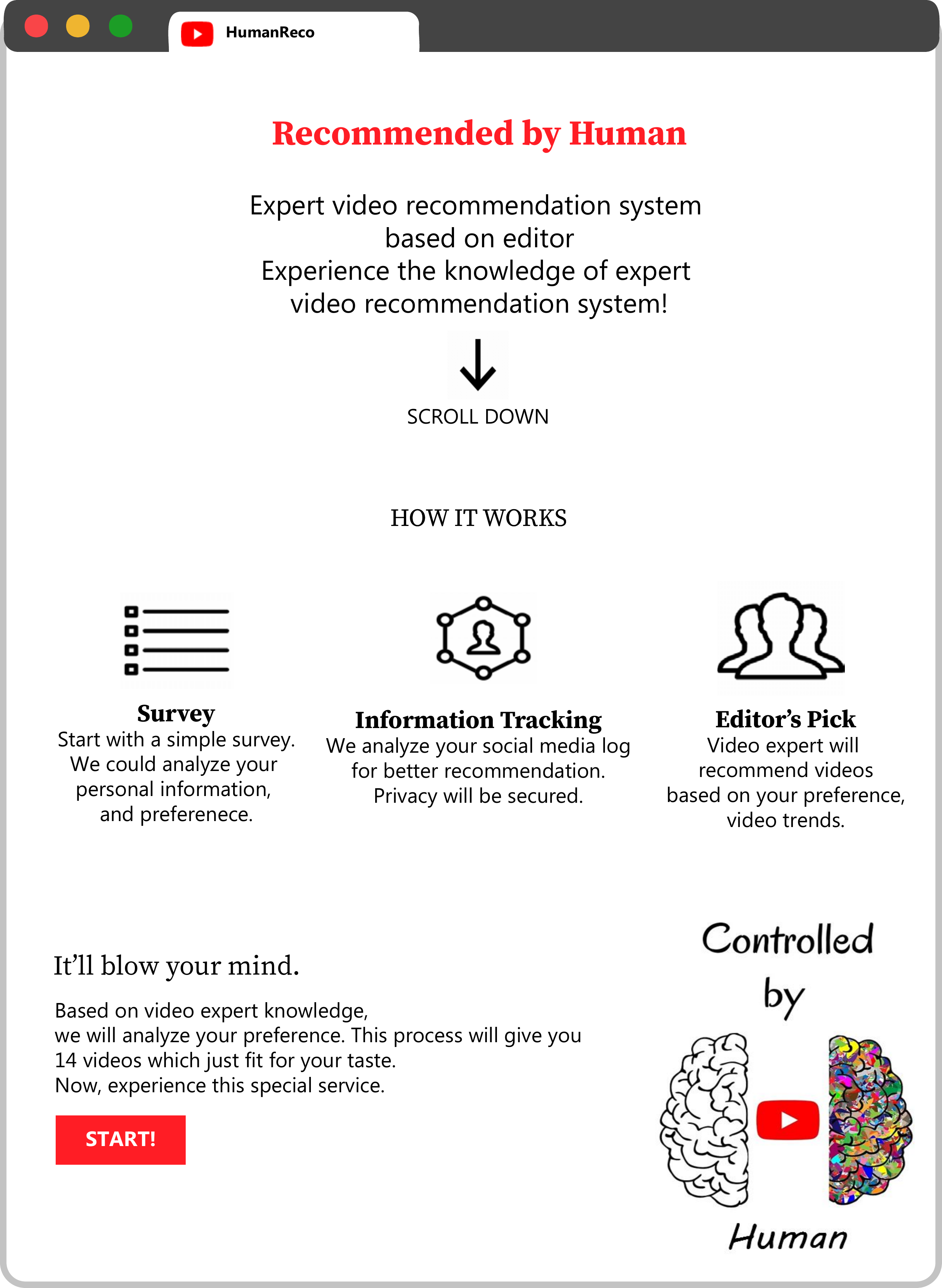}
         \caption{Human condition}
         \label{fig:y equals x}
     \end{subfigure}
  \caption{Introduction web-pages for each condition}
  \label{fig:intro}
\end{figure*}

    We designed a web-based recommendation system by adapting the YouTube interface as in Figure~\ref{fig:intro}. Different recommender identity cues were employed for human and machine conditions~\cite{koh2010heuristic, kim2016interacting}. For the human condition, we proposed the recommendation is executed by an expert video content editor. While we described that artificial intelligence recommends the videos in the machine condition. Details of the whole procedure of the recommendation in each condition are as follows. On the introductory page, participants in the human condition were provided an explanation that human editors recommended the videos, while those in the machine condition were informed that the videos were recommended by machine editors. Participants in the human condition were shown a human editor's image on the recommendation loading page and saw the message that the human editor took part in the recommendation procedures. In the machine condition, participants saw a machine editor's image and received the notification that the algorithm provided the recommendations. Moreover, pre-recorded voices were presented as cues for recommender identities, such that participants in the human condition were exposed to a human voice saying \textit{``a video specialist is in the process of generating a list of videos for you''}. Participants in the machine condition were exposed to a synthetic voice saying \textit{``an algorithm specialized in video recommendations is in the process of generating a list of videos for the participant''}. Female voices, which were neutral and monotone in style, were used in both conditions.

    Participants were then instructed to answer questions regarding demographic information which have no effect on the recommendation performance. This step's purpose was to increase credibility of the recommendation system by inducing each participant to believe that the system uses demographic data to analyze user preferences and provide personalized recommendations. An identical list of recommended videos was given to both human and machine conditions, and seven out of the 14 videos in the recommended list contained inappropriate content. The videos' inappropriateness was assessed by the video violation guideline in YouTube\footnote{\url{https://creatoracademy.youtube.com/page/course/community-guidelines}}. The inappropriate videos included content related to assault, hate speech, fake news, pranks with a threat, and how-to guides of illegal activities (e.g., how to unlock a bicycle chain and how to make guns with a 3D printer). A manipulation check question (``Please list down the inappropriate videos you remember'') was employed to check the inappropriateness of the videos during the pilot test with 15 participants. Six of the fifteen respondents indicated that they saw severely inappropriate videos in the recommended list. Finally, the participants were debriefed about the study’s purpose upon completing the experiment.
    
    The entire procedure of the experiment was as follows:
    
    \begin{enumerate}[label=(\roman*)]
        \item Each participant was instructed to read all descriptions of the recommender on the introductory web page and click the start button at the bottom of the page. 
        \item A survey on users' demographic information was given: name, e-mail address, nationality, age, gender, video interest, and social media user ID were asked.
        \item The participant was advised to wait for the video recommendations as they continued watching the loading page. The loading page contained visual and auditory cues for triggering users to be aware of the recommender identity.
        \item An identical list of pre-selected videos was suggested to the participants. Depending on the conditions, either an image of a machine or a human was placed on the top of the page.
        \item The participant was given two minutes to click and watch the recommended videos on their own choices.
        \item The participant received the confirmation codes to verify that they followed the whole recommendation process to participate in the original survey.   
        \item Upon completion of the survey, the participants were debriefed about the study's purpose. They were made aware that the video list presented in the experiment was not recommended by the real system and the information that they inputted was not saved.

    \end{enumerate}

\subsection{Measures}
    All questionnaire items were measured using a 7-point Likert scale (1 = strongly disagree; 7 = strongly agree). 
  
\subsubsection{Credibility}
    A credibility was measured using four items: 1) The recommendation system has expertise in recommending videos, 2) The recommendation system knows well about the video contents, 3) The recommendation is reliably qualified, and 4) The recommendation system is inexperienced (reversed). The Cronbach's alpha for all four items was found to be 0.89 and these items are validated by prior online service study~\cite{featherman2010reducing}.

\subsubsection{Trust in machine}
    A trust in machine was measured using three items: 1) Machines are more objective than human when performing a task, 2) Machines make better decisions than humans when it comes to ethics, 3) The decision of a machine does not cause damage to others. The Cronbach's alpha for three items was found to be 0.56 and these items are validated by prior machine heuristic research~\cite{sundar2019machine}.

\subsubsection{Dependent variable}
    A dependent variable, attitude toward the recommendation system, was assessed by four items: 1) I am satisfied with the recommendation system, 2) Using this recommendation system is useful, 3) Using this recommendation system is a good idea, 4) It would be uncomfortable if I can't use this recommendation system. The Cronbach's alpha for these items rated 0.92 and validated by prior public perception study~\cite{kim2016interacting}.

\section{Results}
    \subsection{Manipulation check}
    To determine the effectiveness of manipulation in an experimental stimulus, we used different questionnaires in each condition as follows. For the machine condition, participants were instructed to respond to the following agreement statement by 7-point Likert scale: 1) ``\textit{I had a video recommendation from a machine editor}'' (\textit{M} = 4.73, \textit{SD} = 1.38). 
    
    For the human condition, participants responded to the following statement: 2) ``\textit{I had a video recommendation from a human editor}'' (\textit{M} = 4.16, \textit{SD} = 1.29). The participants in machine condition were found to be more likely to recognize the source of the recommendation compared to the human condition (\textit{t}(87) = 2.00, \textit{p} = .047). 

    To explore the participants' engagement by using a 7-point Likert scale in the experiment~\cite{koh2010heuristic}, they were instructed to respond to the following items: 1) ``\textit{How important was the recommendation task for you?}'' 2) ``\textit{How involved were you in the experiment?}'' 3) ``\textit{Did you feel as though you were experiencing an actual recommendation system?}''. No significant difference between machine (\textit{M} = 3.89, \textit{SD} = 1.18) and human conditions (\textit{M} = 3.48, \textit{SD} = 1.37) was observed. Participants in both conditions were found to have been equally motivated to participate in the experiment (\textit{t}(87) = 1.49, \textit{p} = .14).

    \subsection{Hypothesis Tests}
    An independent \textit{t}-test was conducted to address the research question. The result showed the significant difference across attitude toward recommendation system (\textit{t}(87) = 1.99, \textit{p} $<$ .05). Specifically, the participants in the machine condition (\textit{M} = 4.18, \textit{SD} = 1.35) were more likely to have a positive attitude toward the recommended videos than those in the human condition (\textit{M} = 3.62, \textit{SD} = 1.26). Besides, the machine cue triggers the participants' positive attitudes toward recommendations ($\beta$ = 1.06, \textit{p} $<$ .05), but the human cue which indicates the video expert does not affect to the participants' attitudes toward the recommendations (\textit{p} = .86).
    
\begin{figure*}
  \centering
    \begin{subfigure}[b]{0.45\textwidth}
         \centering
         \includegraphics[width=\textwidth]{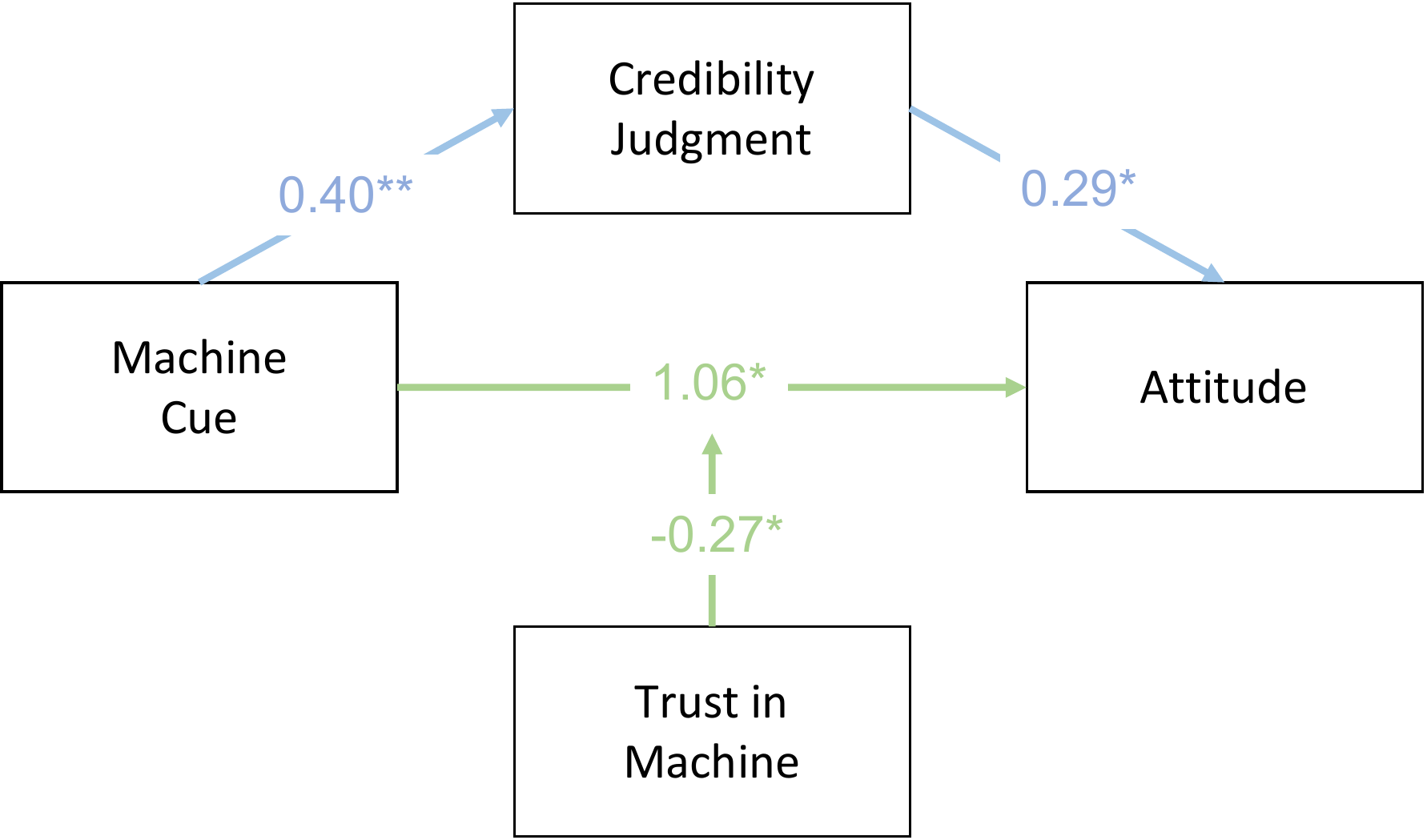}
         \caption{Machine condition}
         \label{fig:machine_result}
     \end{subfigure}
    \quad\quad
    \begin{subfigure}[b]{0.45\textwidth}
         \centering
         \includegraphics[width=\textwidth]{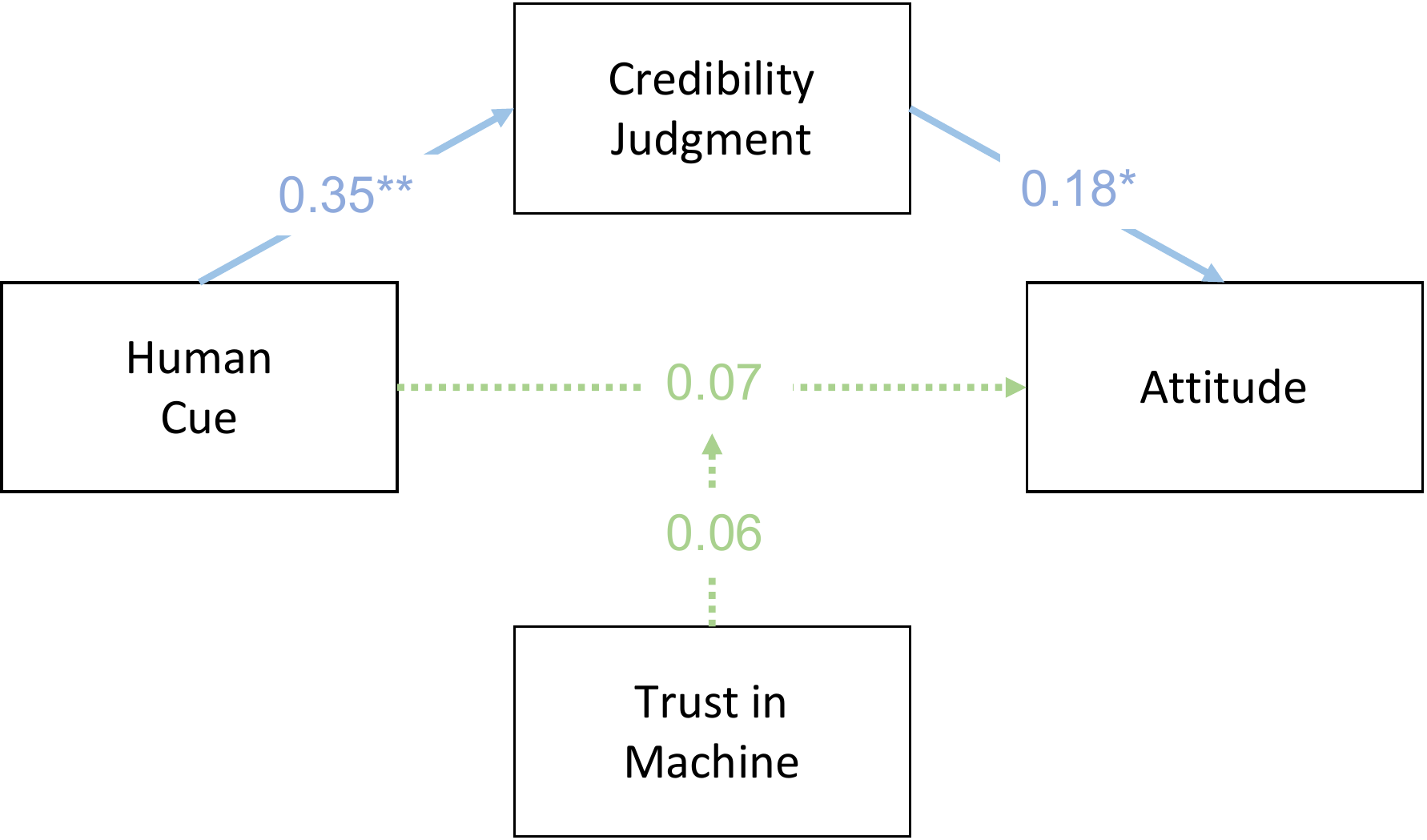}
         \caption{Human condition}
         \label{fig:human_result}
     \end{subfigure}
  \caption{Results of the research models from each condition: machine and human recommender systems (*\textit{p} $<$ 0.05; **\textit{p} $<$ 0.01)}
  \label{fig:result}
\end{figure*}

    Then, we conducted mediation and moderation analysis by using Hayes's PROCESS macro (Model 5) with 5,000 bootstrap samples to answer the hypotheses that we suggested~\cite{hayes2017introduction}. The model was bias-corrected and set at a 95\% confidence interval to analyze each machine's proposed research model and the human condition. All control variables were considered as co-variants. The results are summarized in Figure~\ref{fig:result}.

 
    To validate the H1, which addresses the mediating role of credibility, a simple regression analysis was conducted. The results show that there is a direct effect of machine cue on credibility (\textit{SE} = 0.11, $\beta$ = 0.40, \textit{p} $<$ .01, \textit{CI} = [0.16, 0.64]). That is, machine cue triggers participants to perceive the recommendation system as credible. Besides, the increased credibility dependent on the machine cue leads to have positive attitude toward recommendations (\textit{SE} = 0.11, $\beta$ = .29, \textit{p} $<$ .01, \textit{CI} = [0.09, 0.53]). The human cue also had a direct effect on credibility (\textit{SE} = 0.35, $\beta$ = 0.35, \textit{p} $<$ .01, \textit{CI} = [0.10, 0.60]), where human cue makes the recommendation system more credible. Moreover, as machine condition, increased credibility contributes to positive attitude toward human recommender's recommendations (\textit{SE} = 0.11, $\beta$ = .18, \textit{p} $<$ .01, \textit{CI} = [0.01, 0.46]). Regardless of the interface cue, the higher the credibility, the greater the positive attitude toward recommendations. H2 hypothesized that a positive attitude toward recommendations will be presented among the participants with stronger trust in machines in the presence of machine cues in the interface, rather than a human agent. Results show the significant effect of trust in machine on attitude in presence of machine cue (\textit{SE} = 0.10, $\beta$ = -.27, \textit{p} $<$ .05, \textit{CI} = [-0.48, -0.06]), while there's no significant effect of the trust in machine on the relationship between human cue and attitude (\textit{p} = .59). 

\begin{figure*}[h]
    \centering
    \includegraphics[width=0.8\textwidth]{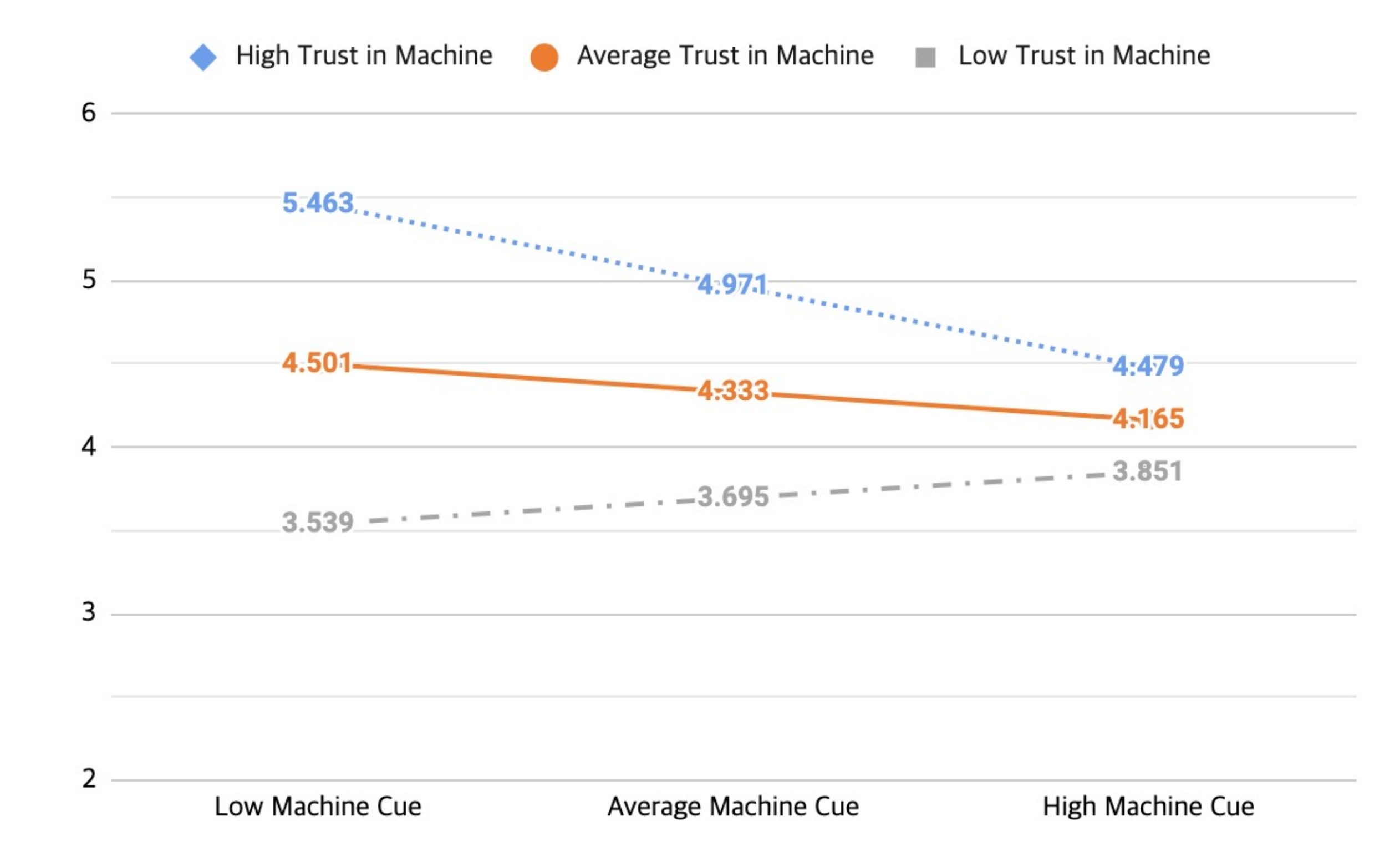}
    \caption{An interaction plot; each line indicates trust level in a machine, x-axis as a degree of perceiving the recommendation system as a machine, and y-axis as an attitude toward algorithmic recommendation system.}
    \label{fig:interact}
\end{figure*}

   Meanwhile, the effect of trust in machines on the relationship between machine cues and attitude is presented in an opposite direction from our expectation. That is, the participant with stronger trust in a machine who saw machine cue has a negative attitude toward recommendations. Therefore, we conducted further analysis to investigate the conditional effects of the variable. The results showed that as trust in the machine and perception that the machine as a main source of the recommendation increased, the attitude becomes more negative (\textit{SE} = 0.10, $\beta$ = -0.35, \textit{p} $<$ .01, \textit{CI} = [-0.57, -0.13]). However, there was no significant relationship between attitude and low (\textit{SE} = 0.14, $\beta$ = 0.11, \textit{p} = .44, \textit{CI} = [-0.18, 0.40]) or average ($\beta$ = -0.12, \textit{p} = .19, \textit{CI} = [-0.30, 0.06]) trust in machine. The interaction points are plotted in Figure~\ref{fig:interact}. Participants with low trust in machines tend to have a positive attitude toward the recommended videos when they are more likely to perceive the recommendation system as a machine. For participants with a more than average level of trust in a machine, the increased level of perceiving the recommendation system as a machine led to a negative attitude toward the recommended videos. That is, the negative attitude can be explained by the participants' dissatisfaction with the recommendations. The participants who have high trust in the machine would expect the performance of the machine recommender system to accurately predict their video preferences. However, as the recommendation list was created regardless of the preferences of the participants, the dissatisfaction might negatively affect the attitude.


\section{Conclusion}

    This study contributes to providing an overview of users' actual experiences when facing a similar situation by conducting the experimental study. As one of the empirical pieces of evidence in addressing cognitive heuristic approaches to online recommendation systems, this study presents the effects of machine-related interface cues on users. In specific, our result from the web-based experiment confirm that people are inclined to exhibit a more positive attitude toward machine-generated recommendations than human-generated recommendations. Even the recommendation video list in our experiment contains videos that violate the content guideline, the participants show cognitive heuristics in terms of revealing trust and reliance in automated suggestions. Besides, the higher the users perceive the recommendation as a machine, the users are more likely to have positive attitudes. From these results, we confirmed that the machine heuristic is involved in building a positive attitude to the recommended videos by algorithms. That is, merely identifying the artificial intelligence as the informant of the interactions can trigger the heuristic based on stereotypes about the operations of machines~\cite{sundar2020rise}. Such automation bias induces people to adhere to automated decision authorities even if the demands seem to violate the common sense of what is right~\cite{mosier1996automation}. With automated cues, people would not verify the given recommendations due to over-credence on the recommendation algorithm. To make matters worse, our results show that regardless of whether an agent is human or automated, the higher credibility of the online system leads to a more positive attitude toward the system. That being said, an authority heuristic could occur by using a well-known recommendation system's interface, YouTube~\cite{sundar2008main}. As the web interface generated for the experiment reveals itself as having an official authority, the participants may confer believability to the recommended content with high credibility to a recommendation system.


    Therefore, the participants would unconsciously consume inappropriate content due to the combined effect of both heuristics. That is, machine cue and its credibility assigned from authority could be the factors that contribute toward spreading misleading and harmful content by making people believe the recommendation performance without in-depth consideration~\cite{papadamou2020pseudoscientific}. In this regard, users always should be aware of the recommended contents rather than their dependence on the various interface cues including algorithmic agent and its credibility.

    While contrary to the findings of previous studies which suggested that the users' greater trust in the machine leads to a higher positive attitude toward the performance of the machine~\cite{sundar2019machine}, the results showed contrasting results. The greater level of users' trust in machines leads to a more negative attitude toward the recommendations in the presence of machine cues. This can be explained by users' experiences of algorithmic operations and general digital literacy; machine cues may have set high expectations for users on online recommender systems~\cite{mosier1996automation}. However, when a user realizes that a specific algorithm fails in a specific task, the cueing machine heuristic can result in negative reactions, such as algorithm aversion~\cite{dietvorst2015algorithm}. Hence, participants with a firmer trust in a machine would have expected the machine to predict the videos they were interested in with some degree of accuracy. Since the video list had been made in advance, the interests of the participants were not considered. After receiving the recommendation, they might have experienced negative disconfirmation between their expectation of the machine's performance and experienced performance, which led to negative attitudes.

    Based on the result, we emphasize the necessity of expanding a research model to explore interactions between humans and machines in a wide range of forecasting domains. In addition to cognitive heuristics to explain humans' preference in using algorithms in prediction tasks, other factors such as algorithm aversion can be considered to explore why people are dissatisfied with the predictions at the same time~\cite{dietvorst2015algorithm}. In other words, multiple factors that trigger interactions in a different direction can be considered simultaneously.

\section{Limitations}  
    Although some academic findings have been presented, several limitations should be addressed in future studies. First, the potential biases regarding the recommendation system might affect the attitude. As human-based video recommendation systems are not usual, users' perceived familiarity with machine recommenders may have positively affected attitudes. This can be assumed by the manipulation check result where the participants in human condition were less likely to recognize the source of recommendation compared to the machine condition. Even though the current study employed the same well-known interface for both conditions, familiarity could have positively affected machine recommenders' perspectives. Besides, the discrepancy between the experiment and the real situation might influence the perceptions of the recommendation system. To increase credibility by giving the sense of personalized recommendation, we inserted made-up stages that ask for personal information including social media id which is not a common procedure in video recommendation platforms. Therefore, researchers should consider a more natural way of making experimental stimuli with high credibility.

\bibliographystyle{unsrt}  
\bibliography{references}

\end{document}